\documentstyle[editedvolume]{crckapb} 
\input psfig.sty

\begin{opening}
\title{PLANETARY NEBULAE:\protect\\
       ABUNDANCES AND ABUNDANCE GRADIENTS}

\author{W.J. MACIEL}
\institute{IAG/USP\\S\~ao Paulo, Brazil}
\end{opening}

\runningtitle{ABUNDANCES OF PLANETARY NEBULAE}

\begin{document}

% The \begin{document} command comes after the \end{opening}
% command.

\paragraph{to appear in:} Chemical evolution of the Milky Way: 
stars versus clusters, ed. F. Giovannelli, F. Matteucci, Kluwer (2000), 
in press.
\bigskip

\begin{abstract}
In this work, a review is given of some recent results and problems 
involved in the determination of chemical abundances of galactic
planetary nebulae, particularly regarding disk and bulge objects.
\end{abstract}

\section{Introduction}

Chemical evolution models for the Galaxy can be included in one of 
four main classes: (i) analytical models, (ii) classical models,
(iii) multiphase models and (iv) chemodynamical models (see for 
example Pagel 1997, Matteucci 1996). These models share a common  
characteristic, that is, they have to satisfy a series of 
observational constraints, most of which are related to the 
determination of chemical abundances. Some examples are the
average gas metallicity in the disk, the age-metallicity relation
and the existence of abundance gradients.

Planetary nebulae (PN) have an important role in the establishment
of these constraints. As it is well known, PN are formed from
the late evolutionary stages of intermediate mass stars ($1 \leq
M/M_\odot \leq 8$), and their chemical composition reflects the
nucleosynthetical processes that occur in these stars. Elements
such as He  and N can be significantly enhanced by the stellar
chemical evolution, while S and Ar are probably unaltered during the
evolution of the progenitor star, so that the nebular abundances of
these elements can be taken as the interstellar abundances at the 
time when the progenitor star was formed.

In the framework of the Peimbert (1978) classification scheme PN can 
be classified as: Type I (disk objects with massive progenitors), 
Type II (disk objects with average mass progenitors), Type III (thick 
disk objects, kinematically detached), Type IV (halo objects), and
Type V (bulge objects) (see also Maciel 1989, Torres-Peimbert and 
Peimbert 1997). 

In this work, some recent results concerning the abundance determinations
of PN are considered, both regarding the abundances themselves
and their spatial and temporal variations. The following sections
discuss PN of types I--III (section 2), type IV (section 3) and 
type V (section 4).

\section{The galactic disk}

\subsection{ABUNDANCES}

Several independent groups have obtained abundances of disk PN, mainly 
from spectroscopic observations of visible and ultraviolet emission
lines coupled with ionization correction factors or
photoionization models. Some recent results are given by Kwitter 
and Henry (1998), Hajian et al. (1997) and Costa et al. (1996a).

Typical errors of the line intensities are of the order of
10--20\%, leading to similar uncertainties in the derived electron
temperatures. The abundances themselves have generally errors
of 0.1--0.2 dex for the best measured elements, such as O/H and 
S/H, and somewhat higher for Ar/H and Cl/H. Geometrical 
effects as well as temperature fluctuations may increase these 
figures, especially for PN with massive central stars (Gruenwald 
and Viegas 1998). A discussion on the accuracy of the derived 
chemical abundances based on hydrodynamical models has been 
recently given by Perinotto et al. (1998).

Table 1 shows average values of the abundances of disk PN  (Costa 
et al. 1996a). Type II has been subdivided into types IIa and IIb, 
according to Fa\'undez-Abans and Maciel (1987), and the table 
also lists types IV and V PN, which will be commented upon later. 
Only the best observed elements are given, though some recent 
work includes less abundant elements such as Si, Mg, Na and Fe 
(Perinotto et al. 1999, Pottasch and Beintema 1999).

\begin{table}[htb]
\begin{center}
\caption{}
\begin{tabular}{lllllll}
\hline
 & I & IIa & IIb & III & IV & V \\
\hline
He/H             & 0.138 & 0.106 & 0.104 & 0.099 & 0.104 & 0.104 \\ 
$\epsilon$(O/H)  & 8.68  & 8.78  & 8.58  & 8.42  & 8.08  & 8.71  \\
$\epsilon$(N/H)  & 8.57  & 8.29  & 7.78  & 7.74  & 7.41  & 8.16  \\
$\epsilon$(S/H)  & 7.04  & 7.02  & 6.83  & 6.74  & 5.64  & 6.87  \\
$\epsilon$(C/H)  & 8.67  & 8.78  & 8.73  & 8.48  & 8.54  &       \\
$\epsilon$(Ne/H) & 8.03  & 8.06  & 7.87  & 7.71  & 7.27  &       \\
$\epsilon$(Ar/H) & 6.61  & 6.47  & 6.26  & 6.07  & 5.22  & 6.22  \\
$\epsilon$(Cl/H) & 5.43  & 5.32  & 5.00  & 4.99  &       & 5.05  \\
                 &       &       &       &       &       &       \\
$\log$(N/O)  & $-$0.11 & $-$0.49 & $-$0.80 & $-$0.68 & $-$0.67 & $-$0.55\\   
$\log$(Ne/O)     & $-$0.65 & $-$0.72 & $-$0.71 & $-$0.71 & $-$0.81 & \\   
$\log$(S/O)  & $-$1.64 & $-$1.76 & $-$1.75 & $-$1.68 & $-$2.44 & $-$1.84 \\   
$\log$(Ar/O) & $-$2.07 & $-$2.31 & $-$2.32 & $-$2.35 & $-$2.86 & $-$2.49 \\   
$\log$(Cl/O)     & $-$3.25 & $-$3.46 & $-$3.58 & $-$3.43 &  & $-$3.66 \\   
                 &       &       &       &       &       &       \\
$z$ (pc)         & 150   & 280   & 420   & 660   & 7200  &       \\
$\Delta v$ (km/s)& 20.5  & 21.3  & 22.1  & 64.0  & 172.8 &       \\
\hline
\end{tabular}
\end{center}
\end{table}

The first group of abundances gives $\epsilon$(X/H) $= \log $(X/H) + 12,
where X/H is the element abundance relative to hydrogen by number 
of atoms. For helium, the table gives the He/H ratio directly. The 
last two rows of the table give the average height $z$ above the 
galactic plane (pc) and the peculiar velocity $\Delta v$ (km/s). It can 
be seen that the He/H ratio increases in the disk along the sequence
III--II--I, similarly to N/H, S/H, Ar/H and Cl/H. For O/H and Ne/H, 
there is an increase from  type III to type II, but the average
abundances of type I PN are not clearly higher than for type II,
which may be partially due to ON cycling in the progenitor stars.

The second group of abundances in Table 1  shows abundances 
relative to oxygen of those elements (Ne, S, Ar, Cl) that are not
produced during the evolution of the progenitor stars, so that 
they can be considered as representative of the interstellar
medium at the time of formation of these stars. Except for halo
(type IV) PN, all objects have essentially constant ratios, a 
result that has been confirmed for S, Ne and Ar for HII regions 
both galactic and extragalactic (Henry and Worthey 1999). The 
ratio N/O is also included, as it is used to distinguish 
between type I and non-type I nebulae. 

Abundance correlations are particularly important in 
order to understand the nucleosynthetical processes ocurring in the 
progenitor stars, since they do not depend on the often uncertain 
distances. Examples are the S/H $\times$ O/H and Ne/H $\times$ O/H  
as well as  N/O $\times$ He/H correlations, which can in principle 
be used to separate the different PN types according to the mass of 
the progenitor star (Costa et al. 1996a, Henry 1998). 

A potentially interesting source of information on PN abundances
is their morphology, particularly since the publication
of detailed high resolution images by Schwarz et al. (1992),
Manchado et al. (1996) and G\'orny et al. (1999). Recent results 
based on HST observations show a clear evidence of N/O enhancements 
in bipolar nebulae (class B), and a corresponding underabundance for 
round PN (class R) (Stanghellini et al. 1999, Stanghellini 1999). 
On the other hand, elliptical PN (class E), which form the majority 
of galactic PN, show a wider range of N/O abundances. The immediate 
conclusion is that B nebulae have relatively massive progenitors, in 
agreement with their being generally closer to the galactic plane, 
while R nebulae are ejected from stars near the lower bracket of the 
intermediate mass stars. E nebulae are probably formed by stars in 
the whole mass interval. As an example, the abundances of 13 bipolar 
nebulae studied by Perinotto and Corradi (1998) agree very well with 
those of type~I objects given in Table~1. Previous evolutionary models 
for AGB stars did not predict the He and N enhancements observed in 
these objects, a situation that has recently changed with the extension 
of the theoretical calculations to stars with $M \simeq 6 M_\odot$ with 
overshooting (Marigo, this conference).

Abundance variations inside the nebula may be important, and are 
presently poorly known (Pottasch 1997). A recent study of bipolar 
nebulae (Perinotto and Corradi 1998) suggests that their observed 
sample is chemically homogeneous.

\subsection{ABUNDANCE GRADIENTS}

The study of radial abundance gradients includes basically (i) the 
average magnitudes of the gradients, (ii) the possible change of slope 
along the galactic disk, and (iii) the time variation of the 
gradients. The first item is relatively well established (cf. Maciel 
1996, 1997, Henry and Worthey 1999), and it seems clear that an 
average gradient of $-0.05$ to $-0.07$ dex/kpc can be observed in the 
Galaxy for O/H, S/H, Ne/H and Ar/H. The PN derived gradient (Maciel
and Quireza 1999, Maciel and K\"oppen 1994, Amnuel 1993, Pasquali and
Perinotto 1993, K\"oppen et al. 1991) is close to -- and slightly 
lower than -- the well known gradient observed from HII regions in 
the Galaxy (Shaver et al. 1983, Afflerbach et al. 1997) and in other 
spiral galaxies (Kennicutt and Garnett 1996, Ferguson et al. 1998). 
Recent work on open cluster stars confirms  these results (Friel 1999,
Phelps, this conference), and also data on B stars, as recently 
discussed  by Smartt and Rolleston (1997), Gummersbach et al. (1998), 
and Smartt (this conference), in contradiction with  earlier work on 
these objects, which reported essentially flat gradients. The radial 
variations of O/H, S/H, Ne/H and Ar/H are consistent with essentially 
constant ratios of S/O, Ne/O and Ar/O from PN and HII regions, as can 
be seen from Table~1 and from a recent discussion on the gradients 
in the Galaxy and in other galaxies (Henry and Worthey 1999).

\begin{figure}
%
%----------------------------------------------------------------
\centerline{\psfig{figure=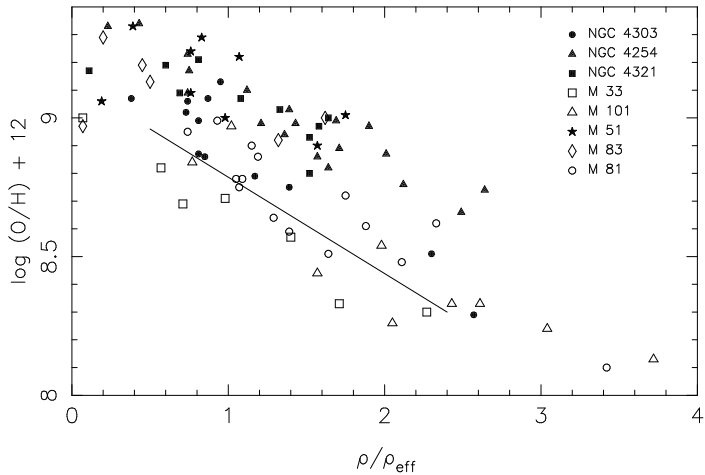,height=6.5 cm,angle=270} }
\caption{O/H radial gradient in spiral galaxies and PN (straight
   line).}
%----------------------------------------------------------------
%
\end{figure}

As an illustration, figure 1 shows the O/H gradient derived from HII 
regions in several spiral galaxies (Henry et al. 1994), along with 
the average gradient for disk PN in the Galaxy (straight line) 
from Maciel and Quireza (1999). In this figure, the abscissa gives the 
distance to the centre of the galaxy in terms of the effective radius, 
defined as the radius where half of the optical emission is contained 
(cf. Maciel 1997). The similarity of the PN and HII region gradients 
suggests that orbital diffusion is probably mild in the radial direction, 
since the PN population is at least a few Gyr older than the HII regions.
Azimuthal diffusion could be much higher, however, as the PN are 
sampled all over the galactic disk. Therefore, figure~1 suggests that 
the gas is homogeneously distributed in the azimuthal direction. This 
can be confirmed by plotting the O/H abundances of PN in a concentric 
ring of width $\Delta R$ around the galactic centre as a function of the 
longitude. No correlations are observed, even if relatively wide rings  
($\Delta R \simeq 1-2$ kpc) are considered. On the other hand, according 
to Wielen et al. (1996) the gradients are not much affected by orbital 
diffusion provided the surface density  of stars falls off exponentially 
with the galactocentric distance $R$, and both the gradient and the 
diffusion coefficient do not appreciably change with~$R$.

\begin{figure}
%
%----------------------------------------------------------------
\centerline{\psfig{figure=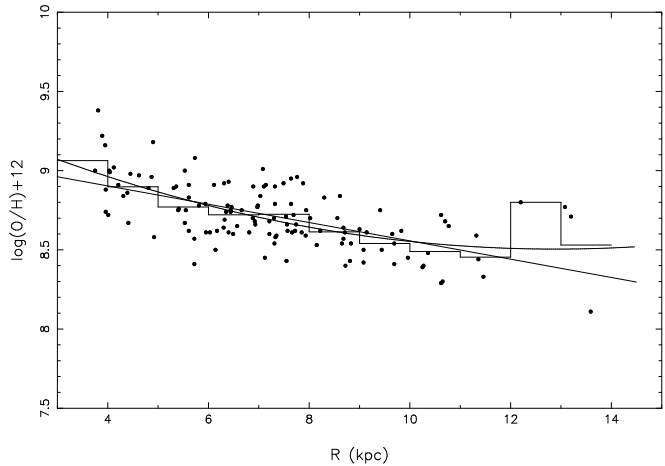,height=6.5 cm,angle=270} }
\caption{O/H gradient from disk PN (Maciel and Quireza 1999).}
%----------------------------------------------------------------
%
\end{figure}

Radial variations of the gradients of O/H, S/H, Ne/H and Ar/H have
been studied by Maciel and Quireza (1999). As an example, figure~2
shows the O/H gradient, where some recent data for planetary nebulae 
in the outer galactic disk from Costa et al. (1997) have been included.
The dots are disk PN; the straight line  gives a linear fit corresponding 
to $-0.058$ dex/kpc; the curve is a second-order  polynomial fit 
corresponding to a gradient of $-0.058$ dex/kpc at the sun, where 
$R_0 = 7.6 $ kpc, and the step function has been fitted with 1 kpc bins. 
A flattening at large radial distances can be observed, in agreement 
with some recent work on HII regions (V\'\i lchez and Esteban 1996).

The time variation of the abundance gradients is not as well defined, 
basically due to the fact that the ages attributed to PN are not 
sufficiently well known and the age difference between HII regions
and B stars is relatively small. Type III PN are generally older and 
show flatter gradients, but part of this may be due to orbital diffusion. 
Including HII regions, B stars  and the different types of PN, the 
present results are consistent with some steepening of the  gradients, 
in agreement with earlier  suggestions by Maciel and  K\"oppen (1994), 
Golovaty and Malkov (1997) and Maciel and Quireza  (1999). The latter 
obtained a rough estimate of the average steepening rate as $-0.004$ 
dex kpc$^{-1}$ Gyr$^{-1}$. Despite the large uncertainty of this 
result, it clearly  shows that any time variation in the gradients  
is probably small, so that the possibility of a constant gradient 
cannot be ruled out. 

Radial gradients and their variations are extremely important as 
constraints to chemical evolution models. Recent classical  models 
predict similar gradients as observed in the framework of an inside-out 
scenario, showing some flattening near the outer Galaxy and a time 
steepening (Chiappini et al. 1997, Matteucci and Chiappini 1999). 
Chemodynamical models (Samland et al. 1997, Hensler 1999) also predict 
some flattening at large $R$, and time steepening gradients are predicted 
by theoretical models by G\"otz and K\"oppen (1992) and  Tosi (1988). 
On the other hand, multiphase models (Ferrini et al. 1994, Moll\'a et 
al. 1997)  suggest just the opposite behaviour, as also models developed 
by Prantzos and  Silk (1998), Boissier and  Prantzos (1999) and Allen 
et al. (1998), so that new independent observations and estimates 
of  the gradients are needed to settle this question.

\section{The halo}

Only an handful of halo PN are known, and in some cases it is not clear 
whether the nebula is a true halo object or belongs to the thick disk. 
These PN have very low metal abundances as can be seen in Table~1, 
which shows very clearly halo--disk variations ranging from 0.10 dex 
to an order of magnitude, therefore much higher than the expected 
abundance uncertainties. A recent analysis of the halo object DDDM1 
(Kwitter and Henry 1998) shows that the He, N, C, O, and Ne abundances 
are similar to those of type IV PN in Table~1. These abundances suggest 
that type IV PN reflect the metal poor conditions in the halo at the 
time of the formation of their progenitor stars, displaying a halo--disk 
vertical  gradient, as found years ago by Fa\'undez-Abans and Maciel 
(1988) and confirmed by Cuisinier (1993), in contrast with an essentially 
flat vertical gradient for disk PN (Fa\'undez-Abans and Maciel 1988, 
Pasquali and Perinotto 1993, Cuisinier 1993).

From the second set of abundances of Table~1, it can be seen that
halo PN have much lower S/O, Ne/O and Ar/O ratios than the remaining 
types, apart from being more metal-poor. According to Henry and Worthey 
(1999), these discrepancies could be caused by local abundance 
fluctuations due to unmixed ejecta from recent supernova events.

An additional problem is that several of the PN associated with the 
halo and thick disk (types III and IV) have probably low mass progenitors 
with strong stellar winds (Maciel et al. 1990, Maciel 1993). In fact, 
some of these nebulae apparently have central stars with masses under 
0.55~$M_\odot$, in contradiction with most evolutionary tracks 
for intermediate mass stars. 

\section{The bulge}

Many bulge, or type V PN (Maciel 1989) are known (see for example 
Beaulieu et al. 1999), but only recently accurate abundances  
have been obtained (Ratag et al. 1997, Costa et al. 1996b). Recent 
work by Cuisinier et al. (1999) and Costa and Maciel (1999) has led
to He, O, N, Ar and S abundances to about 40 bulge PN, with an 
uncertainty comparable to disk objects. However, van Hoof
and van de Steene (1999) have found a larger scatter among the 
recent determinations for a sample of 5 bulge PN, which was
attributed to uncertainties in the electron temperature.

A comparison of He/H and N/O abundances in the bulge and in the 
disk shows that objects with higher ratios are less present in the 
bulge, suggesting that it contains an excess of older progenitor stars, 
since younger, massive objects are generally overabundant in these 
elements. On the other hand, the bulge metal abundances are  generally 
comparable with those of the disk, and the O/H, Ar/H and S/H ratios can 
be higher than the disk counterparts even though very metal rich 
PN are missing in the bulge. Since underabundant nebulae are also 
present, these results suggest that the bulge contains a mixed population, 
so that star formation in the bulge probably spans a wide time interval. 

Chemical abundances of PN in the bulge of M31 have been recently
studied by Jacoby and Ciardullo (1999), who have included in
their analysis some results by Stasi\'nska et al. (1998) and
Richer et al. (1999). Figure~3 shows a comparison of this sample 
(top panel) with the galactic bulge objects from Ratag et al. (1997) 
(lower panel, thin line) and Cuisinier et al. (1999)
(lower panel, thick line). It can ben seen that the O/H abundance 
distributions are very similar, peaking around 8.7 dex, and showing 
very few if any super metal rich objects with supersolar abundances. 

\begin{figure}
%
%----------------------------------------------------------------
\centerline{\psfig{figure=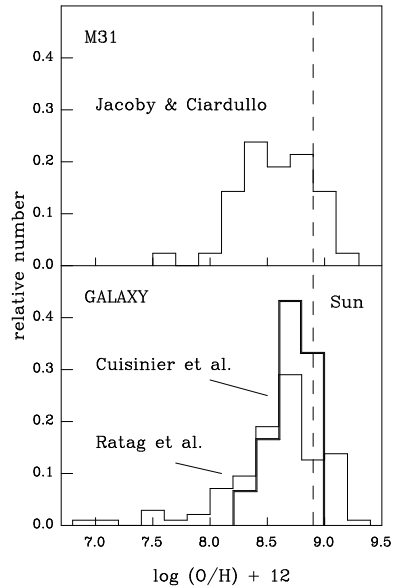,height=8.0 cm,angle=0} }
\caption{O/H distributions of bulge PN in the Galaxy and in M31.}
%----------------------------------------------------------------
\end{figure}

This fact has been interpreted by Jacoby and Ciardullo (1999) as an 
inability of metal rich stars to produce sufficiently bright PN and 
to a possible gradient in the bulge of M31. Coupled with the He/H and 
N/O abundances, it would suggest again that the youngest and most metal 
rich PN are less frequent in the bulge.

It is interesting to compare the PN metallicity distribution with the
[Fe/H] distribution of stars in the bulge. A direct comparison depends
on the adopted [O/Fe] $\times$ [Fe/H] relationship, as Fe abundances 
are known for a limited sample of PN only. Adopting the [O/Fe] $\times$ 
[Fe/H] relation for the {\it solar neighbourhood} (Matteucci et al. 1999), 
it can be seen from figure~4 that the observed peak in the PN 
metallicity distribution for the bulge (dotted line) looks similar to 
the distribution of K giants in Baade's Window (McWilliam and Rich 1994, 
dashed line), but depleted of the very metal rich objects. In this case, 
the peak of the distribution occurs for [Fe/H] close to zero, which is 
also similar to the Mira variables metallicity distribution shown by 
Feast (this conference). Taking into account the predicted relationship 
of Matteucci et al. (1999) for the {\it bulge}, which assumes a faster 
evolution of the bulge compared to the halo, the observed peak of the PN
metallicity distribution is displaced by about 0.5 dex towards lower 
metallicities (solid line), as in this case a given [Fe/H] metallicity 
implies a larger [O/Fe] ratio. As a result, the derived distribution 
shows an even stronger depletion of the very metal rich objects.

\begin{figure}
%
%----------------------------------------------------------------
\centerline{\psfig{figure=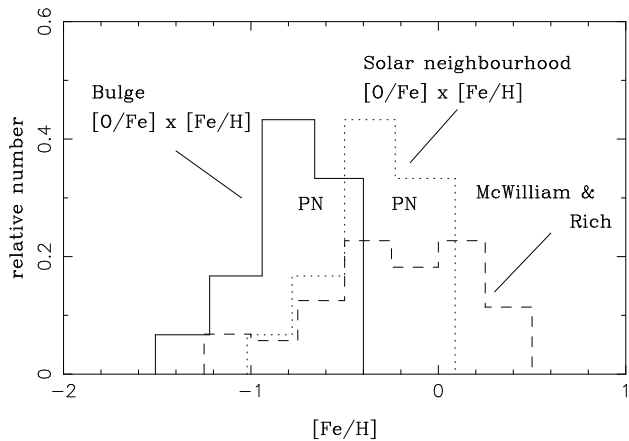,height=6.0 cm,angle=270} }
\caption{Metallicity distribution of bulge PN and K giants.}
%----------------------------------------------------------------
\end{figure}

Several reasons can be considered to explain the differences in
the metallicity distributions of figure~4, namely (i) systematic errors 
in the PN abundances, (ii) ON cycling in the PN progenitor stars, 
(iii) statistical uncertainties, and (iv) uncertainties in the 
adopted [O/Fe] $\times$ [Fe/H] relationships. As discussed by Maciel 
(1999), the first three are probably not important enough to account 
for a 0.5 dex discrepancy as shown in figure~4, so that we suggest that 
the main reason is an excess of oxygen enhancement relative to iron in the
bulge [O/Fe] $\times$ [Fe/H] relation from Matteucci et al. (1999).
This is supported by recent results by Barbuy (1999) for the bulge
globular cluster NGC 6528 and for the six giants in the McWilliam
and Rich (1994) sample for which [OI] lines have been measured.

\smallskip\noindent
{\small\it Acknowledgements. \small This work was partially 
supported by CNPq and FAPESP}.

\bigskip\noindent
{\bf References}
\medskip
{\small
\par\noindent
Afflerbach, A.,  Churchwell, E., Werner, M.W.: 1997, {\it ApJ \bf 478}, 190
\par\noindent
Allen, C.,  Carigi, L., Peimbert, M.: 1998, {\it ApJ \bf 494}, 247
\par\noindent
Amnuel, P.R.: 1993, {\it MNRAS \bf 261}, 263
\par\noindent
Barbuy, B.: 1999, {\it Galaxy evolution}, ed. M. Spite, F. Crifo, Kluwer 
    (in press)
\par\noindent
Beaulieu, S.F., Dopita, M.A., Freeman, K.C.: 1999, {\it ApJ \bf 515}, 610
\par\noindent
Boissier, S., Prantzos, N.: 1999, {\it MNRAS \bf 307}, 857
\par\noindent
Chiappini, C., Matteucci, F., Gratton, R.: 1997, {\it ApJ \bf 477}, 765
\par\noindent
Costa, R.D.D., Chiappini, C., Maciel, W. J., Freitas-Pacheco, J.A.: 1996a, 
    {\it A\&AS \bf 116}, 249
\par\noindent
Costa, R.D.D., Freitas-Pacheco, J.A., Fran\c ca, J.A.: 1996b, {\it A\&A 
    \bf 313}, 924
\par\noindent
Costa, R.D.D., Chiappini, C., Maciel, W. J., Freitas-Pacheco, J.A.: 1997, 
    {\it Advances in stellar evolution}, ed. R.T. Rood, 
    A. Renzini, Cambridge, 159
\par\noindent
Costa, R.D.D., Maciel, W.J.: 1999, {\it Galaxy evolution}, ed. M. Spite, 
    F. Crifo, Kluwer (in press) 
\par\noindent
Cousinier, F.: 1993, {\it Acta Astron. \bf 43}, 455
\par\noindent
Cousinier, F., Maciel, W.J., K\"oppen, J., Acker, A., Stenholm, B.:  
   1999 {\it A\&A} (in press)
\par\noindent
Fa\'undez-Abans, M., Maciel, W.J.: 1987, {\it A\&A \bf 183}, 324
\par\noindent
Fa\'undez-Abans, M., Maciel, W.J.: 1988, {\it Rev. Mex. A \& A \bf 16}, 105
\par\noindent
Ferrini, F., Moll\'a, M., Pardi, C., D\'\i az, A.I.: 1994, 
    {\it ApJ \bf 427}, 745
\par\noindent
Ferguson, A.M.N., Gallagher, J.S., Wyse, R.F.G.: 1998, {\it  AJ 
    \bf 116}, 673
\par\noindent
Friel, E.D.: 1999, {\it Galaxy evolution}, ed. M. Spite, F. Crifo, Kluwer
    (in press) 
\par\noindent
G\"otz, M., K\"oppen, J.: 1992, {\it A\&A \bf 262}, 455
\par\noindent
Golovaty, V.V., Malkov, Yu.F.: 1997, {\it IAU Symp. 180}, ed. H.J. 
    Habing, H.J.G.L.M. Lamers, Kluwer, 408
\par\noindent
G\'orny, S.K., Schwarz, H.E., Corradi, R.L.M., van Winckel, H.: 1999, 
    {\it A\&AS \bf 136}, 145
\par\noindent
Gruenwald, R.B., Viegas, S.M.M.: 1998, {\it ApJ \bf 501}, 221
\par\noindent
Gummersbach, C.A., Kaufer, A., Schaefer, D.R., Szeifert, T., Wolf, B.: 
    1998, {\it A\&A \bf 338}, 881
\par\noindent
Hajian, A.R., Balick, B., Terzian, Y., Perinotto, M.: 1997, {\it ApJ \bf 
    487}, 313
\par\noindent
Henry, R.B.C.: 1998, {\it  ASP Conference Series vol. 147}, ed. 
    D. Friedli, M. Edmunds, C. Robert, L. Drissen, 59
\par\noindent
Henry, R.B.C., Pagel, B.E.J., Chincarini, G.L.: 1994, {\it MNRAS \bf 
    266}, 421 
\par\noindent
Henry, R.B.C., Worthey, G.: 1999, {\it  PASP \bf 111}, 919
\par\noindent
Hensler, G.: 1999, {\it Galaxy evolution}, ed. M. Spite, F. Crifo, Kluwer
 (in press)
\par\noindent
Jacoby, G.H., Ciardullo, R.: 1999, {\it ApJ \bf 515}, 169
\par\noindent
Kennicutt, R.C., Garnett, D.R.: 1996, {\it  ApJ \bf 456}, 504
\par\noindent
K\"oppen, J., Acker, A., Stenholm, B.: 1991, {\it A\&A \bf 248}, 197
\par\noindent
Kwitter, K.B., Henry, R.B.C.: 1998, {\it  ApJ \bf 493}, 247
\par\noindent
Maciel, W.J.: 1989, {\it IAU Symp. 131}, ed. S. Torres-Peimbert, 
    Kluwer, 73
\par\noindent
Maciel, W.J.: 1993, {\it A\&SS \bf 209}, 65
\par\noindent
Maciel, W.J.: 1996, {\it Stellar abundances}, ed. B. Barbuy, W.J. Maciel,
    J.C.Greg\'orio-Hetem, IAG/USP, 79
\par\noindent
Maciel, W.J.: 1997, {\it IAU Symp. 180}, ed. H.J. Habing, H.J.G.L.M.
    Lamers, Kluwer, 397
\par\noindent
Maciel, W.J.: 1999, {\it A\&A } (in press)
\par\noindent
Maciel, W.J., Freitas-Pacheco, J.A., Codina, S.J.: 1990, {\it A\&A 
    \bf 239}, 301
\par\noindent
Maciel, W.J., K\"oppen, J.: 1994, {\it A\&A \bf 282}, 436
\par\noindent
Maciel, W.J., Quireza, C.: 1999, {\it A\&A \bf 345}, 629
\par\noindent
Manchado, A., Guerrero, M., Stanghellini, L., Serra-Ricart, M.: 1996, 
    {\it The IAC Morphological Catalog of Northern Galactic Planetary 
    Nebulae}, IAC
\par\noindent
Matteucci, F., 1996: {\it Fund. Cosm. Phys. \bf 17}, 283
\par\noindent
Matteucci, F., Chiappini, C.: 1999, {\it Chemical evolution from zero to
   high redshift}, ed. J.R. Walsh, M.R. Rosa, ESO (in press)
\par\noindent
Matteucci, F., Romano, D., Molaro, P.: 1999, {\it A\&A \bf 341}, 458
\par\noindent
McWilliam, A., Rich, R.M.: 1994, {\it ApJS \bf 91}, 749
\par\noindent
Moll\'a, M., Ferrini, F., D\'\i az, A.I.: 1997, {\it  ApJ \bf 75}, 519
\par\noindent
Pagel, B.E.J.: 1997, {\it Nucleosynthesis and chemical evolution
    of galaxies}, Cambridge
\par\noindent
Pasquali, A., Perinotto, M.: 1993, {\it A\&A \bf 280}, 581
\par\noindent
Peimbert, M.: 1978, {\it IAU Symp. 76}, ed. Y. Terzian, Reidel, 215
\par\noindent
Perinotto, M., Bencini, C.G., Pasquali, A., Manchado, A., Rodriguez
    Espinoza, J.M., Stanga, R.: 1999, {\it A\&A \bf 347}, 967
\par\noindent
Perinotto, M., Corradi, R.L.M.: 1998, {\it A\&A \bf 332}, 721
\par\noindent
Perinotto, M., Kifonidis, K., Sch\"onberner, D., Marten, H.: 1998, 
     {\it A\&A \bf 332}, 1044
\par\noindent
Pottasch, S.R.: 1997, {\it IAU Symp. 180}, ed. H.J. Habing, H.J.G.L.M. 
    Lamers, Kluwer, 483
\par\noindent
Pottasch, S.R., Beintema, D.A.: 1999, {\it A\&A \bf 347}, 975
\par\noindent
Prantzos, N., Silk, J.: 1998, {\it ApJ \bf 507}, 229
\par\noindent
Ratag, M.A., Pottasch, S.R., Dennefeld, M., Menzies, J.W.: 1997, {\it 
     A\&AS \bf 126}, 297
\par\noindent
Richer, M.G., Stasinska, G., McCall, M.L.: 1999, {\it A\&AS \bf 135}, 203
\par\noindent
Samland, M., Hensler, G., Theis, C.: 1997, {\it ApJ \bf 476}, 544
\par\noindent
Schwarz, H.E., Corradi, R.L.M., Melnick, J.: 1992, {\it A\&AS \bf 96}, 23
\par\noindent
Shaver, P.A., McGee, R.X., Newton, L.M., Danks, A.C., Pottasch, S.R.: 
    1983, {\it MNRAS \bf 204}, 53
\par\noindent
Smartt, S.J., Rolleston, W.R.J.: 1997, {\it ApJ \bf 481}, L47
\par\noindent
Stanghellini, L.: 1999, {\it Ap\&SS} (in press, astro-ph/9906188) 
\par\noindent
Stanghellini, L., Blades, C.J., Osmer, S.J., Barlow, M.J., Liu, X.W.: 
    1999, {\it ApJ \bf 510}, 687
\par\noindent
Stasinska, G., Richer, M.G., McCall, M.L.: 1998, {\it A\&A \bf 336}, 667
\par\noindent
Torres-Peimbert, S., Peimbert, M.: 1997, {\it IAU Symp. 180}, ed. H.J. 
    Habing, H.J.G.L.M. Lamers, Kluwer, 175
\par\noindent
Tosi, M.: 1988, {\it A\&A \bf 197}, 47
\par\noindent
van Hoof, P.A.M., van de Steene, G.C.: 1999, {\it MNRAS \bf 308}, 623
\par\noindent
V\'\i lchez J.M., Esteban, C.: 1996, {\it MNRAS \bf 280}, 720
\par\noindent
Wielen, R., Fuchs, B., Dettbarn, C.: 1996, {\it A\&A \bf 314}, 438

\end{document}